\documentclass{ptephy_v1}

\preprintnumber{XXXX-XXXX} 

\usepackage{graphicx}
\usepackage{dcolumn}
\usepackage{bm}

\usepackage{ulem}

\usepackage{color}
\input{colordvi.tex}

\newcommand{\beqa}{\begin{eqnarray}}
\newcommand{\eeqa}{\end{eqnarray}}

\newcommand{\vp}{\varphi}
\newcommand{\barR}{\overline{R}}
\newcommand{\barg}{\overline{g}}
\newcommand{\barnabla}{\overline{\nabla}}
\newcommand{\barH}{\overline{H}}

\newcommand{\barp}{\overline{p}}

\begin{document}

\title{Cosmological constraints on scalar-tensor gravity 
and the variation of the gravitational constant}

\author{Junpei Ooba}
\affil{Department of physics and astrophysics, Nagoya University, Nagoya 464-8602, Japan}

\author{Kiyotomo Ichiki${}^{1,}$}
\affil{Kobayashi-Maskawa Institute for the Origin of Particles and the Universe, Nagoya University, Nagoya 464-8602, Japan}

\author{Takeshi Chiba}
\affil{Department of Physics, College of Humanities and Sciences, Nihon University, Tokyo 156-8550, Japan}

\author{Naoshi Sugiyama${}^{1,2,}$}
\affil{Kavli Institute for the Physics and Mathematics of the Universe (Kavli IPMU), The University of Tokyo, Chiba 277-8582, Japan}


\begin{abstract}
We present cosmological constraints on the scalar-tensor theory of gravity
by analyzing the angular power spectrum data of the cosmic microwave background
obtained from the Planck 2015 results together with  the baryon acoustic oscillations (BAO) data. 
We find that the inclusion of the BAO data improves the constraints
on the time variation of the effective gravitational constant by more than 10\%, that is, 
the time variation of the effective gravitational constant between the recombination and 
the present epochs is constrained as
$G_{\rm rec}/G_0-1 <1.9\times 10^{-3}\ (95.45\%\ {\rm C.L.})$ and 
$G_{\rm rec}/G_0-1 <5.5\times 10^{-3}\ (99.99 \%\ {\rm C.L.})$.
We also discuss the dependence of  the constraints on the choice of the prior. 
\end{abstract}

\subjectindex{E03, E50, E60, E63}

\maketitle


\section{\label{sec:level1}Introduction}

The existence of scalar fields whose vacuum expectation values determine
 the physical constants is generically predicted by the recent attempts
 toward unifying all elementary forces in nature based on string theory
 \cite{string}.  In this context, scalar-tensor theories of gravity are
 a natural alternative to the Einstein gravity since they arise from
 the low-energy limit of string theory.  In the scalar-tensor theories
 of gravity, a scalar field couples to the Ricci scalar, which provides
 a natural framework for realizing the time variation of the
 gravitational constant via the dynamics of the scalar field.  In the
 Jordan-Brans-Dicke theory of gravity \cite{BDT,BDT2}, which is the simplest
 example of scalar-tensor theories, a constant coupling parameter
 $\omega$ is introduced.  In more general scalar-tensor theories
 \cite{st,st2,st3}, $\omega$ is promoted to a function of the Brans-Dicke scalar
 field $\phi$.  In the limit $\omega \rightarrow \infty$, the Einstein
 gravity is recovered and the gravitational constant becomes a constant
 in time.

The coupling parameter $\omega$ has been constrained by several solar
system experiments.  For instance, the weak-field experiment conducted
in the Solar System by the Cassini mission has put strong constraints on
the post-Newtonian deviation from the Einstein gravity, where $\omega$
is constrained as $\omega>43000$ at a $2\ \sigma$ level \cite{Cassini,Will}.

On cosmological scales, the possibility of constraining
the Brans-Dicke theory by temperature and polarization anisotropies of
the cosmic microwave background (CMB) was suggested in \cite{ck}, and
Nagata {\it et al.} \cite{NCS2004} first placed constraints on a general
scalar-tensor theory called the harmonic attractor model including the
Jordan-Brans-Dicke theory \cite{dn,dn2}.  In this model the scalar field has
a quadratic effective potential of positive curvature in the Einstein
frame, and the Einstein gravity is an attractor that naturally
suppresses any deviations from the Einstein gravity in the present
epoch.  Nagata {\it et al.} reported that the present-day value of
$\omega$ is constrained as $\omega>1000$ at a $2\ \sigma$ level by analyzing
the CMB data from the Wilkinson Microwave Anisotropy Probe
(WMAP). Moreover, the gravitational constant at the recombination epoch
$G_{\rm rec}$ relative to the present gravitational constant $G_0$ is
constrained as $G_{\rm rec}/G_0<1.05$ ($2\ \sigma$). These constraints
basically come from the fact that the size of the sound horizon at the
recombination epoch, which determines the characteristic angular scale
in the angular power spectrum of CMB anisotropies, depends on the
amounts of matter and baryon contents and on the strength of gravity
at that epoch.
Recently, we have analyzed the CMB power spectra data from Planck 2015  \cite{Planck} 
in the harmonic attractor model to put constraints on the deviations from general relativity \cite{ooba}.
We find 
a constraint on $\omega$ as $\omega>2000$ at $95.45\%$ confidence level (C.L.), and
an order-of-magnitude improvement on the change of $G$: $G_{\rm rec}/G_0<1.0056 \ (1.0115)$ 
at $95.45\%$ C.L. ( $99.99\%$ C.L.) \cite{ooba}--- see also
\cite{Avilez} for the analysis in the Brans-Dicke gravity (a constant $\omega$) model and
\cite{Umilta,Ballardini} for the analysis in the induced gravity dark energy model.

Acoustic peaks in the CMB power spectrum are transferred to peaks in
baryons through the coupling between photons and baryons through the
Thomson scattering, and these acoustic peaks are later imprinted on the
matter power spectrum; they are known as baryon acoustic oscillations
(BAO).  BAO have been measured by a number of galaxy redshift
surveys. Since the BAO measurements are basically geometrical, like CMB
acoustic peaks, they can be used to break parameter degeneracies in the
analysis based solely on the CMB data.  In this paper, we further
improve the constraints on the scalar-tensor theory by including the
recent measurements of BAO \cite{6dF,boss, sdss}.

The remainder of the paper is organized as follows. In Sect. 2 we
explain the scalar-tensor cosmological model, and we describe our method
for constraining the scalar-tensor coupling parameters in Sect. 3.  In
Sect. 4, we compare the model with the CMB data and BAO data. The prior
dependence of the analysis is also discussed.  We summarize our
results in Sect. 5.

\section{Model}

The action describing a general massless scalar-tensor theory in the 
Jordan frame is given by \cite{NCS2002}
\begin{equation}
S =\frac{1}{16\pi G_0}\int d^4x\sqrt{-g}\left[ \phi R - \frac{\omega(\phi)}{\phi}(\nabla \phi)^2 \right] + S_{\rm m}[\psi,g_{\mu\nu}],
\label{eq:action}
\end{equation}
where $G_0$ is the present-day Newtonian gravitational constant and $S_{\rm
m}[\psi,g_{\mu\nu}]$ is the matter action, which is a function of the
matter variable $\psi$ and the metric $g_{\mu\nu}$.  We regard this
``Jordan frame metric'' as defining the lengths and times actually measured
by laboratory rods and clocks, since in the action Eq. (\ref{eq:action})
matter is universally coupled to $g_{\mu\nu}$ \cite{de,cy}.  The
function $\omega(\phi)$ is the dimensionless coupling parameter, which
depends on the scalar field $\phi$. The deviation from the Einstein
gravity depends on the asymptotic value of $\phi$ at spatial
infinity. According to the cosmological attractor scenario \cite{dn,dn2},
the dynamics of $\phi$ in the Friedmann universe is analogous to that of
a particle attracted toward the minimum of its effective potential
with a friction (the Hubble friction in the Friedmann universe) in the Einstein frame.
The effective potential corresponds
to the logarithm of the conformal factor.  Since a potential near a
minimum is generically parabolic, we study the case where the effective
potential is quadratic.  This setup corresponds to $\omega(\phi)$ of the
following form:
\begin{equation}
2\,\omega(\phi) + 3 = \left\{ {\alpha_0}^2 - \beta\, {\rm ln}(\phi/\phi_0) \right\}^{-1},
\label{eq:omega}
\end{equation}
where $\phi_0$ is the present value of $\phi$ and  $\alpha_0$ and $\beta$ are model parameters.  
See Appendix \ref{app1} for details. 

The background equations for a Friedmann universe are
\begin{equation}
\rho' = -3\frac{a'}{a}(\rho + p),
\label{eq:EoS}
\end{equation}
\begin{equation}
\left( \frac{a'}{a} \right)^2 + K= \frac{8\pi G_0\,\rho\, a^2}{3\, \phi} - \frac{a'}{a}\frac{\phi'}{\phi} + \frac{\omega}{6}\left( \frac{\phi'}{\phi} \right)^2,
\label{eq:friedmann}
\end{equation}
\begin{equation}
\phi'' + 2\frac{a'}{a}\phi' = \frac{1}{2\, \omega + 3}\left\{ 8\pi G_0\,a^2(\rho - 3p) - {\phi'}^2\frac{d\omega}{d\phi} \right\},
\label{eq:phiEoM}
\end{equation}
where $a$ is the cosmological scale factor and the prime notation denotes a derivative with respect to the conformal time,
$\rho$ and $p$ are the total energy density and pressure, respectively,
and $K$ denotes a constant spatial curvature.

The effective gravitational constant measured by Cavendish-type experiments is given by \cite{de}
\begin{equation}
G(\phi) = \frac{G_0}{\phi}\frac{2\, \omega(\phi) + 4}{2\, \omega(\phi) + 3}.
\label{eq:Gphi}
\end{equation}
The present value of $\phi$ must yield the present-day Newtonian gravitational constant and satisfy the expression $G(\phi_0) = G_0$.
Thus, we have
\begin{equation}
\phi_0 = \frac{2\, \omega_0 + 4}{2\, \omega_0 + 3} = 1 + {\alpha_0}^2,
\label{eq:phi0}
\end{equation}
where $\omega_0$ is the present value of $\omega(\phi)$.

\begin{figure}[ht]
\centering\includegraphics[width=9cm,]{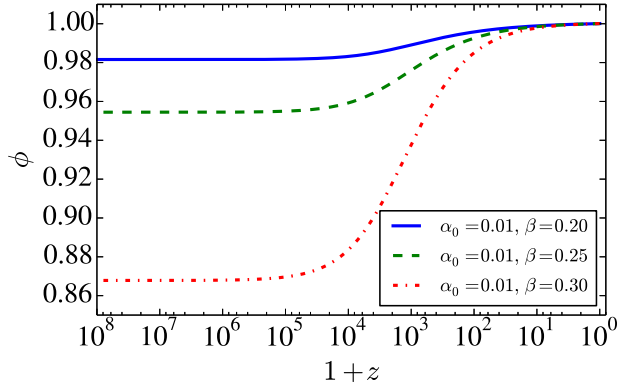}
\caption{\label{fig:BDphi} Time evolution of $\phi$ in the scalar-tensor
 $\Lambda {\rm CDM}$ model, with the parameters as indicated in the
 figure. The other cosmological parameters are fixed to the standard values.}
\end{figure}

\begin{figure}[ht]
\centering\includegraphics[width=9cm,]{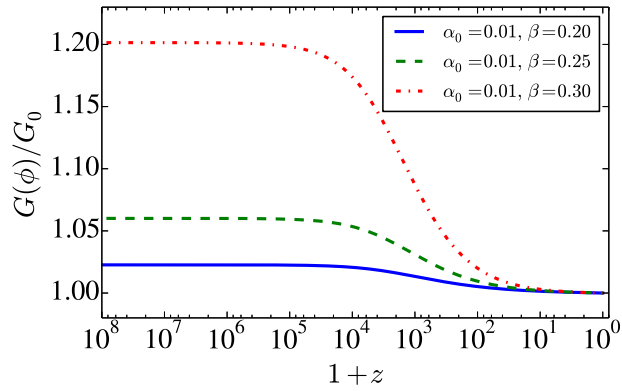}
\caption{\label{fig:Geff} Time evolution of $G(\phi)/G_0$ in the
 scalar-tensor models with the same parameters as in Fig.~\ref{fig:BDphi}.
The effective gravitational constant $G(\phi)$ is inversely proportional to the scalar field $\phi$ through Eq. (\ref{eq:Gphi}).}
\end{figure}

Typical evolutions of $\phi$ and $G(\phi)$ are shown in Figs. \ref{fig:BDphi} and \ref{fig:Geff}, respectively.
Here $h = 0.68$ and $\Omega_{\rm m}h^2 = 0.14$  are assumed, where 
$h$ is the dimensionless Hubble parameter and $\Omega_{\rm m}$ is the matter density parameter. 
In the radiation-dominated epoch, $\phi$ becomes almost constant 
because the pressure of the relativistic component in Eq. (\ref{eq:phiEoM}) is $p=\rho/3$.
After the matter-radiation equality, $\phi$ begins to increase  up to the present value $\phi_0$. 
The variation in the value of $\phi$ alters the Hubble parameter in the early universe
from its value under the Einstein gravity through Eq. (\ref{eq:friedmann}). 
Therefore, we expect that observational data during the matter-dominated 
era, such as  CMB and especially  BAO, are useful in putting constraints on the scalar-tensor gravity.

\begin{figure}[ht]
\centering\includegraphics[width=9cm,]{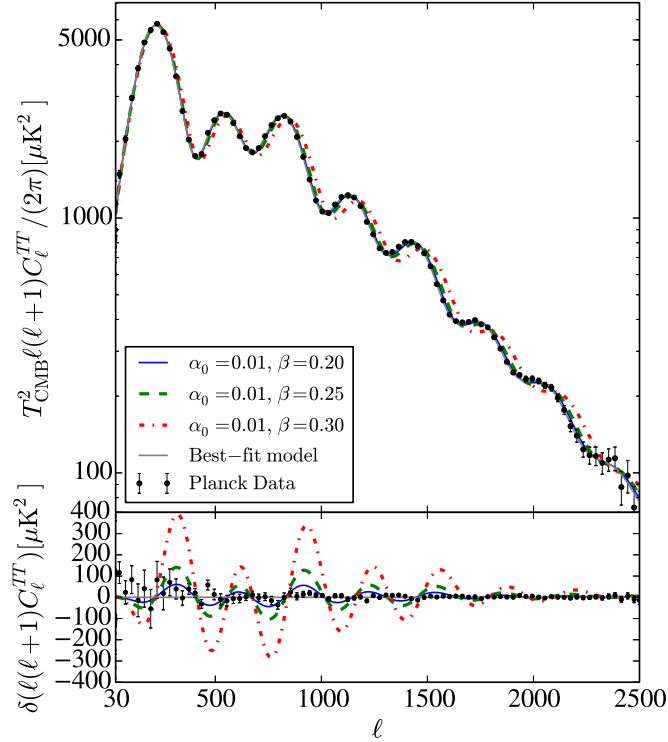}
\caption{\label{fig:errorCl} CMB temperature anisotropy spectra in the scalar-tensor models with the $\Lambda {\rm CDM}$ parameters.
The data points with error bars represent the Planck data.
The gray solid line shows the best-fit $\Lambda{\rm CDM}$ theoretical model fitted to the Planck + BAO data.
Residuals with respect to the best-fit model are shown in the lower panel.}
\end{figure}

Typical CMB temperature anisotrpy spectra are shown in Fig. \ref{fig:errorCl}.
Here, $h = 0.6782$, $\Omega_{\rm b}h^2 = 0.02227$, $\Omega_{\rm c}h^2 = 0.1185$, $\tau_{\rm reio} = 0.067$,
${\rm ln}(10^{10}A_{\rm s}) = 3.064$, $n_{\rm s} = 0.9684$, $T_{\rm CMB} = 2.7255\ \rm K$, $N_{\rm eff} = 3.046$
are assumed for the parameters of the $\Lambda \rm CDM$  model where  
$\Omega_{\rm b}$ and $\Omega_{\rm c}$ are the density parameters for
baryon and cold dark matter components, respectively, $\tau_{\rm reio}$ is
the reionization optical depth, and $A_{\rm s}$ and $n_{\rm s}$ are the amplitude and
spectral index of primordial curvature fluctuations, respectively. 
Since the locations of the acoustic peaks and the damping scale depend differently on the horizon length at recombination,
we can constrain the $\phi$-induced variations in the horizon scale by analyzing the measurements of the CMB anisotropies at small angular scales. 
The positions of the acoustic peaks  are proportional to the horizon length ($\propto H^{-1}$),
while that of the damping scale is less affected by it ($\propto \sqrt{H^{-1}}$).
Therefore, the locations of the first peak and the diffusion tail in the angular power spectrum become closer as the expansion rate becomes larger,
suppressing the small-scale peaks, as shown in Fig. \ref{fig:errorCl}.


\section{Methods}

To compute the temperature and polarization fluctuations in the CMB and the lensing potential power spectra, we
numerically solve the equations in the model described in the previous
section by modifying the publicly available numerical code, CLASS \cite{CLASS2}.  The data are 
 analyzed using
the Markov chain Monte Carlo (MCMC) method with Monte Python \cite{Audren},
developed in the CLASS code.
In our calculations, we consider ($\alpha_0$,\ $\beta$) in Eq. (\ref{eq:omega}), which characterize the scalar-tensor theory,
in addition to the parameters of the $\Lambda {\rm CDM}$ model.

We set the priors for the standard cosmological parameters as
\begin{align}
\label{eq:prior}
H_0 \in (30,100),\ \ \Omega_{\rm b}h^2 \in (0.005,0.04), \nonumber\\
\Omega_{\rm c}h^2 \in (0.01,0.5),\ \ \tau_{\rm reio} \in (0.005,0.5), \\
{\rm ln}(10^{10}A_{\rm s}) \in (0.5,10),\ \ n_{\rm s} \in (0.5,1.5), \nonumber
\end{align}
and for $\alpha_0$ and $\beta$ as
\begin{align}
\label{eq:prior2}
\rm log_{10}(\alpha_0) &\in (-6,-0.5), \\
\beta &\in (0,0.4).
\label{eq:prior3}
\end{align}
The CMB temperature and the effective number of neutrinos were set
to $T_{\rm CMB}= 2.7255\ \rm K$ from COBE \cite{Fixsen} and $N_{\rm
eff}=3.046$, respectively.
The primordial helium fraction $Y_{\rm He}$ is inferred from the standard Big Bang nucleosynthesis, as a function of the baryon density.

\begin{figure}[ht]
\centering\includegraphics[width=9cm,]{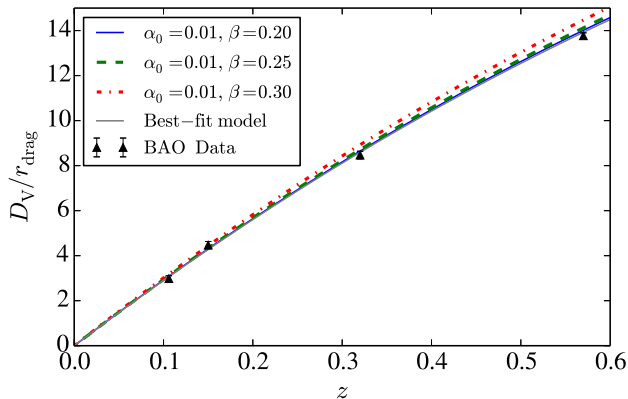}
\caption{\label{fig:BAO} {Time evolutions of $D_{\rm V}/r_{\rm drag}$ in the
 scalar-tensor models with the same parameters as in Fig.~\ref{fig:BDphi}.
 The data points with error bars represent the data of the BAO measurements.
 The gray solid line shows the best-fit $\Lambda{\rm CDM}$ theoretical model fitted to the Planck + BAO data.}
}
\end{figure}

We compare our results with the CMB angular power spectrum data from the
Planck 2015 mission \cite{Planck} and the BAO measurements in the matter power spectra obtained by
the 6dF Galaxy Survey (6dFGS) \cite{6dF}, the Baryon Oscillation Spectroscopic Survey (BOSS; LOWZ and CMASS) \cite{boss},
and the Sloan Digital Sky Survey (SDSS) main galaxy sample (MGS) \cite{sdss}.
The Planck data include the auto power spectra of
temperature and polarization anisotropies ({\it TT} and {\it EE}), their
cross-power spectrum ({\it TE}), and the lensing potential power
spectrum. The data of the BAO measurements are the values of
$D_{\rm V}/r_{\rm drag}$ as shown in Fig. \ref{fig:BAO}, where $r_{\rm
drag}$ is the coming sound horizon at the end of the baryon drag epoch
and $D_{\rm V}$ is the function of the angular diameter distance $D_{\rm
A}(z)$ and Hubble parameter $H(z)$ defined by
\begin{equation}
D_{\rm V}(z) = \left[ (1+z)^2 D_{\rm A}^2(z)\frac{z}{H(z)} \right]^{1/3}.
\label{eq:Dv}
\end{equation}

The BAO can be used to constrain the scalar-tensor cosmological
models as the CMB: the length of the sound horizon at the end of the
baryon drag epoch scales as $r_{\rm drag}\propto H^{-1} (z_{\rm drag})
\propto G^{-1}_{\rm drag}$, while the geometric distance indicator scales
as $D_{\rm V}\propto H^{-1} \propto G^{-1}_{\rm bao}$, where $G_{\rm
drag}$ and $G_{\rm bao}$ are the gravitational constant at the redshifts
of the baryon
drag epoch and the BAO measurements, respectively. Therefore, if $G_{\rm
drag} \neq G_{\rm bao}$, the BAO data can be used to constrain the
scalar-tensor cosmological models. Indeed, the models considered in
this paper always predict $G_{\rm drag}>G_{\rm bao}$, leading to a larger
$D_{\rm V}/r_{\rm drag}$, as is shown in Fig. \ref{fig:BAO}. 

\begin{figure}[ht]
\centering\includegraphics[width=9cm,]{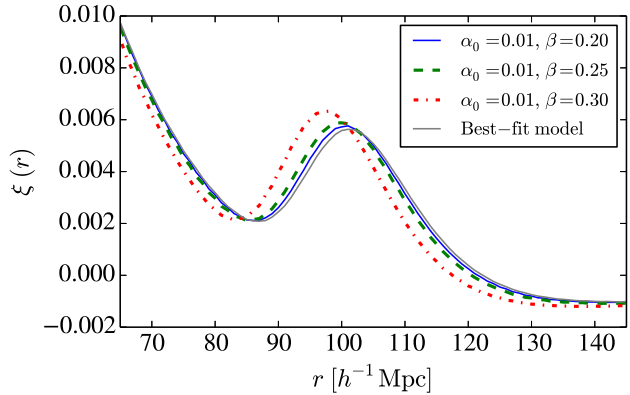}
\caption{\label{fig:xi} {Two point correlation function $\xi (r)$ in the
 scalar-tensor models with the same parameters as in Fig.~\ref{fig:BDphi}.
 The gray solid line shows the best-fit $\Lambda{\rm CDM}$ theoretical model fitted to the Planck + BAO data.}
}
\end{figure}

The two-point correlation function is defined by
\begin{equation}
\xi (r) = \int \frac{k^2dk}{2\pi^2}\frac{\sin(kr)}{kr}P(k),
\label{eq:xi}
\end{equation}
where $r$ is the distance, $k$ is the wave number,
and $P(k)$ is the power spectrum of primordial curvature fluctuations.
Some typical examples in the scalar-tensor model are shown in Fig. \ref{fig:xi}.
The BAO peak scale is proportional to $r_{\rm drag}$,
and therefore the location of the BAO peak moves to smaller scale as the $G_{\rm drag}$ becomes larger.

Because the variation of the gravitational constant could alter the distance to the last scattering surface of the CMB
through the change in the Hubble parameter,
its effect on the angular power spectrum may degenerate with the effects of spatial curvature in the Friedmann universe and
the effective number of relativistic degrees of freedom.
Therefore, we separately perform MCMC analyses for models with the spatial curvature ($\Omega_{\rm K}$) and
with the effective number of relativistic degrees of freedom ($N_{\rm eff}$).
We set the priors for $\Omega_{\rm K}$ and $N_{\rm eff}$ as
\begin{align}
\Omega_{\rm K} &\in (-0.5,0.5), \\
N_{\rm eff} &\in (1,5),
\end{align}
while the same priors are used for the other standard cosmological
parameters and ($\alpha_0$,\ $\beta$) as shown in Eqs. (\ref{eq:prior})--(\ref{eq:prior3}).

\section{\label{sec:level1}Results}

We show the results of the parameter constraints for flat universe models 
(Sect. 4.1), for non-flat universe models (Sect. 4.2) and for models with $N_{\rm eff}$ (Sect. 4.3).

\subsection{\label{sec:level2}Flat universe case} 

\begin{figure}[ht] 
\centering\includegraphics[width=7cm,]{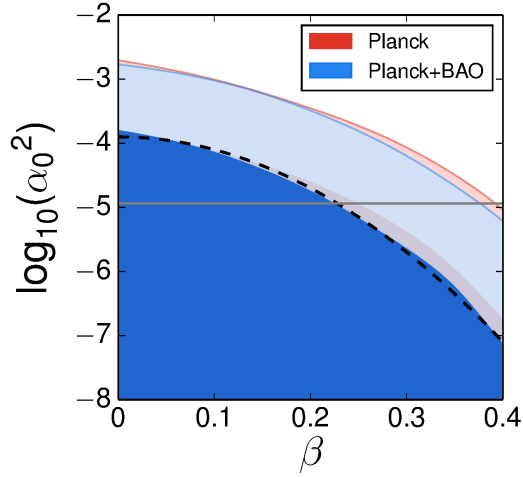}
\caption{\label{fig:flat_BAOcontour} Contours at $95.45\%$ and $99.99\%$ confidence levels
in the $\log_{10}({\alpha_0}^2)$--$\beta$ plane
for the scalar-tensor $\Lambda {\rm CDM}$ models with the other parameters marginalized,
using the Planck data only (red) or the Planck+BAO data (blue). 
The black dashed line shows the function $\log_{10}({\alpha_0}^2) = -3.9-20\beta^2$
and the gray solid line shows the bound from the Solar System experiment.
}
\end{figure}

In Fig. \ref{fig:flat_BAOcontour}, we show the constraint contours in the $\log_{10}({\alpha_0}^2)$--$\beta$ plane, where the other parameters are marginalized.
We find that the constraints on $\log_{10}({\alpha_0}^2)$ and $\beta$ are approximately given by
\begin{align}
\log_{10}({\alpha_0}^2) &< -3.9-20\beta^2\ \ (95.45\%), \\
\log_{10}({\alpha_0}^2) &< -2.8-20\beta^2\ \ (99.99\%),
\end{align}
where the numbers in parentheses denote the confidence level. 
These results can be translated into the present-day value of the coupling parameter $\omega$ at $\beta = 0$ using Eq. (\ref{eq:omega}) as
\begin{align}
\omega &> 3254\ \ (95.45\%), \\
\omega &> 307\ \ (99.99\%).
\end{align}
These limits are little changed compared with those obtained by the Planck data alone:
$\omega > 3224\ (303)$ 
at $95.45\%$ C.L. ( $99.99\%$ C.L.).

Table \ref{tab:table1} shows the $68.27\%$ confidence limits of the standard cosmological parameters
in the scalar-tensor $\Lambda \rm CDM$ model.
These parameters are still consistent with those of the Planck results \cite{Planck} in the standard $\Lambda {\rm CDM}$ model.
Table \ref{tab:table2} shows the $95.45\%$ confidence limits of the parameters ${\rm log}_{10}({\alpha_0}^2)$ and $\beta$.\\

\begin{figure}[ht]
\centering\includegraphics[width=7cm,]{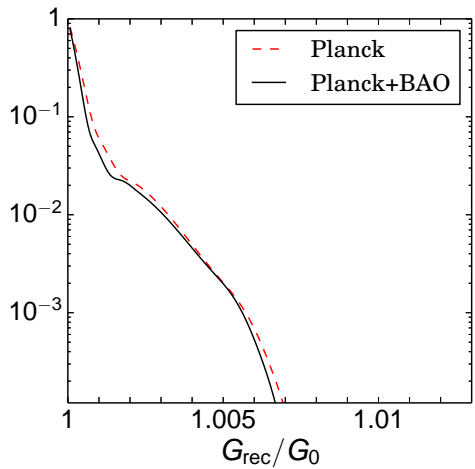}
\caption{\label{fig:flat_BAOGphi_1d} Posterior distribution of $G_{\rm rec}/G_0$,
using the Planck data only (red dashed) or the Planck+BAO data (black).
}
\end{figure}

Next, we consider the variation of the gravitational constant in the recombination epoch.
We define $G_{\rm rec} \equiv G(\phi_{\rm rec})$ and put constraints on $G_{\rm rec}/G_0$,
after marginalizing over the other parameters.
Here, $\phi_{\rm rec}$ is the value of $\phi$ at the recombination epoch when the visibility function takes its maximum value. 
We compute the  marginalized posterior distribution of $G_{\rm rec}/G_0$
as shown in Fig. \ref{fig:flat_BAOGphi_1d} (for flat models).   
We find that $G_{\rm rec}/G_0$ is constrained as
\begin{align}
\label{eq:Gbound1}
G_{\rm rec}/G_0 -1 &< 1.9\times 10^{-3}\ \ (95.45\%), \\
G_{\rm rec}/G_0 -1 &< 5.5\times 10^{-3}\ \ (99.99\%).
\label{eq:Gbound2}
\end{align}
These are 10\% improvements over the results obtained
by the Planck data alone:  $G_{\rm rec}/G_0 - 1 < 2.1\times 10^{-3} \ (6.0\times 10^{-3})$ 
at $95.45\%$ C.L. ( $99.99\%$ C.L.).

\begin{table*}[!h]
\caption{\label{tab:table1}
$68.27\%$ confidence limits for the standard cosmological parameters in 
the scalar-tensor $\Lambda \rm CDM$ model.
}
\centering
\begin{tabular}{lcccr}
\hline
\hline
&\multicolumn{3}{c}{$68.27\%$\ limits}\\
\textrm{Parameter}&
\textrm{$\Omega_{\rm K} = 0$}&
\textrm{$\Omega_{\rm K} \neq 0$}&
\textrm{$N_{\rm eff} \neq {\rm const.}$}\\
\hline
$\Omega_{\rm b}h^2$ & $0.02232\pm 0.00014$ & $0.02225\pm 0.00015$ & $0.02231\pm 0.00019$\\
$\Omega_{\rm c}h^2$ & $0.1183\pm 0.0011$ & $0.1193\pm 0.0014$ & $0.1182\pm 0.0028$\\
$H_0$ & $68.00\pm 0.49$ & $68.53\pm 0.74$ & $67.96\pm 1.17$\\
$\tau_{\rm reio}$ & $0.072\pm 0.012$ & $0.069\pm 0.012$ & $0.072\pm 0.012$\\
${\rm ln}(10^{10}A_{\rm s})$ & $3.074\pm 0.023$ & $3.071\pm 0.023$ & $3.075\pm 0.024$\\
$n_{\rm s}$ & $0.9675\pm 0.0042$ & $0.9651\pm 0.0049$ & $0.9672\pm 0.0071$\\
\hline
$\Omega_{\rm K}$ & --- & $0.0019\pm 0.0020$ & --- \\
$N_{\rm eff}$ & --- & --- & $3.035\pm 0.170$\\
\hline
\hline
\end{tabular}
\end{table*}

\begin{table}[!h]
\caption{\label{tab:table2}
$95.45\%$ confidence limits for ${\rm log}_{10}({\alpha_0}^2)$ and $\beta$.
}
\centering
\begin{tabular}{lcccr}
\hline
\hline
&\multicolumn{3}{c}{$95.45\%$\ limits}\\
\textrm{Parameter}&
\textrm{$\Omega_{\rm K} = 0$}&
\textrm{$\Omega_{\rm K} \neq 0$}&
\textrm{$N_{\rm eff} \neq {\rm const.}$}\\
\hline
${\rm log}_{10}({\alpha_0}^2)$ & $<-4.56$ & $<-4.58$ & $<-4.48$\\
$\beta$ & $<0.418$ & $<0.423$ & $<0.417$\\
\hline
\hline
\end{tabular}
\end{table}

\subsection{\label{sec:level2}Non-flat universe case}

We also perform an MCMC analysis including the spatial curvature parameter $\Omega_{\rm K}$.
This is motivated by the fact that the attractor model used in this paper would predict a larger gravitational constant in the past,
pushing the acoustic peaks toward smaller angular scales.
This effect could be compensated with the positive curvature, which brings back the peaks toward larger angles \cite{NCS2004}.
This degeneracy, however, should be broken using the CMB data on diffusion damping scales,
because the curvature does not affect the diffusion damping whereas the variation of the gravitational constant does, as discussed above.

\begin{figure}[ht] 
\centering\includegraphics[width=7cm,]{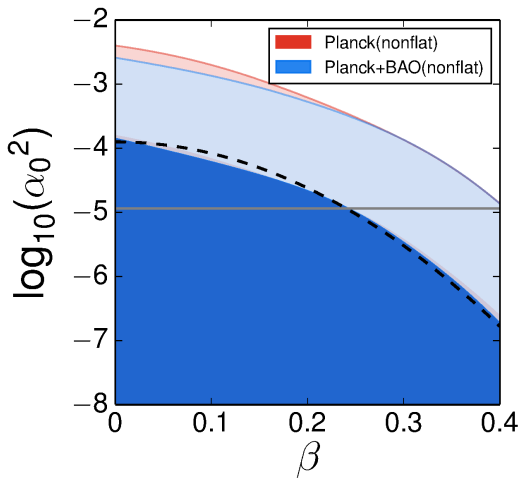}
\caption{\label{fig:nonflat_BAOcontour} Contours for $95.45\%$ and $99.99\%$ confidence levels 
in the $\log_{10}({\alpha_0}^2)$--$\beta$ plane
for the scalar-tensor non-flat $\Lambda {\rm CDM}$ models with the other parameters
 marginalized for the Planck data only (red) or for the Planck+BAO data (blue).
The black dashed line and the gray solid line show the function 
$\log_{10}({\alpha_0}^2) = -3.9-18\beta^2$
and the bound from the Solar System experiment, respectively.
}
\end{figure}

The constraints on the parameters $\log_{10}({\alpha_0}^2)$ and $\beta$ in non-flat models are shown in Fig. \ref{fig:nonflat_BAOcontour},
where the other parameters including $\Omega_{\rm K}$ are marginalized.
We find that the constraints on the scalar-tensor coupling parameters are hardly affected by the inclusion of the spatial curvature.
This is because the angular power spectrum on small angular scales obtained from Planck is 
so precise as to break the degeneracy between the effects of the varying gravitational constant and the spatial curvature.
We find that $\log_{10}({\alpha_0}^2)$ is constrained approximately as
\begin{align}
\log_{10}({\alpha_0}^2) &< -3.9-18\beta^2\ \ (95.45\%), \\
\log_{10}({\alpha_0}^2) &< -2.7-18\beta^2\ \ (99.99\%),
\end{align}
and the coupling parameter $\omega$ as
\begin{align}
\omega &> 3124\ \ (95.45\%), \\
\omega &> 258\ \ (99.99\%).
\end{align}
We find that the inclusion of the spatial curvature does not much affect the constraint at the 95.45\% confidence limit,
while slightly weakens the constraint at the 99.99\% confidence limit.

\begin{figure}[ht] 
\centering\includegraphics[width=7cm,]{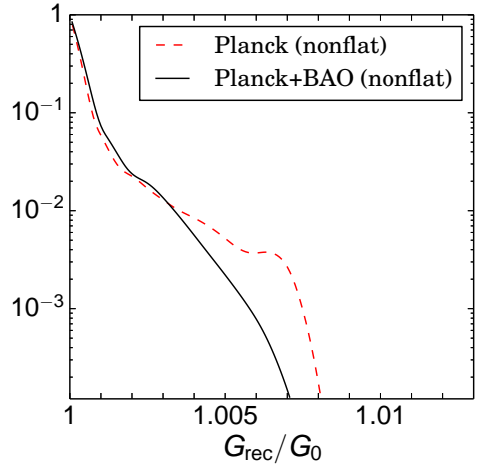}
\caption{\label{fig:nonflat_BAOGphi_1d} Posterior distribution of $G_{\rm rec}/G_0$ for 
the non-flat models,
using the Planck data only (red dashed) or the Planck+BAO data (black).
}
\end{figure}

Also, we find that $G_{\rm rec}/G_0$ in the non-flat universe is constrained as
\begin{align}
G_{\rm rec}/G_0 -1&< 1.9\times 10^{-3} \ \ (95.45\%), \\
G_{\rm rec}/G_0 -1&< 6.2\times 10^{-3} \ \ (99.99\%).
\end{align}
The posterior distribution of $G_{\rm rec}/G_0$ is shown in Fig. \ref{fig:nonflat_BAOGphi_1d}.
The inclusion of the spatial curvature makes only minor changes on the constraints. 
The center column in Table \ref{tab:table1} shows $68.27\%$ confidence limits of the cosmological parameters
in the scalar-tensor non-flat $\Lambda \rm CDM$ model.
These parameters are also still consistent with the those of the Planck results \cite{Planck}.
The limits on ${\rm log}_{10}({\alpha_0}^2)$ and $\beta$ are summarized in Table \ref{tab:table2}.

\subsection{\label{sec:level2}Flat universe case including $N_{\rm eff}$}

\begin{figure}[ht] 
\centering\includegraphics[width=7cm,]{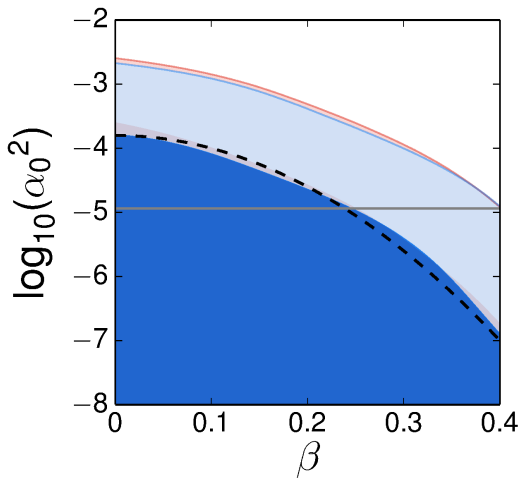}
\caption{\label{fig:Neff_BAOcontour} Contours for $95.45\%$ and $99.99\%$ confidence levels 
in the $\log_{10}({\alpha_0}^2)$--$\beta$ plane
for the scalar-tensor flat $\Lambda {\rm CDM}$ models including $N_{\rm eff}$ with the other parameters
marginalized for the Planck data only (red) or for the Planck+BAO data (blue).
The black dashed line and the gray solid line show the function 
$\log_{10}({\alpha_0}^2) = -3.8-20\beta^2$
and the bound from the Solar System experiment, respectively.
}
\end{figure}

Additionally, we perform an MCMC analysis including the effective number of relativistic degrees of freedom $N_{\rm eff}$.
This is motivated by the fact that the attractor model used in this paper would predict a larger Hubble parameter value in the past.
This effect could be compensated with a smaller $N_{\rm eff}$,
which predicts a smaller Hubble parameter particularly before the recombination epoch, and alters the diffusion damping scale.
This degeneracy, however, should be broken if we consider the distance to the CMB,
because the energy density of radiation components decays away in the matter-dominated era
while the variation of the gravitational constant continues to affect the expansion of the universe during that era.
The constraints on the parameters $\log_{10}({\alpha_0}^2)$ and $\beta$ in models with $N_{\rm eff}$ are shown in Fig. \ref{fig:Neff_BAOcontour},
where the other parameters including $N_{\rm eff}$ are marginalized.
We find that the constraints on the scalar-tensor coupling parameters are slightly affected by the inclusion of $N_{\rm eff}$.
We find that $\log_{10}({\alpha_0}^2)$ is constrained approximately as
\begin{align}
\log_{10}({\alpha_0}^2) &< -3.8-20\beta^2\ \ (95.45\%), \\
\log_{10}({\alpha_0}^2) &< -2.6-20\beta^2\ \ (99.99\%),
\end{align}
and the coupling parameter $\omega$ as
\begin{align}
\omega &> 2917\ \ (95.45\%), \\
\omega &> 177\ \ (99.99\%).
\end{align}
We find that the inclusion of $N_{\rm eff}$ slightly weakens the constraints. 

\begin{figure}[ht] 
\centering\includegraphics[width=7cm,]{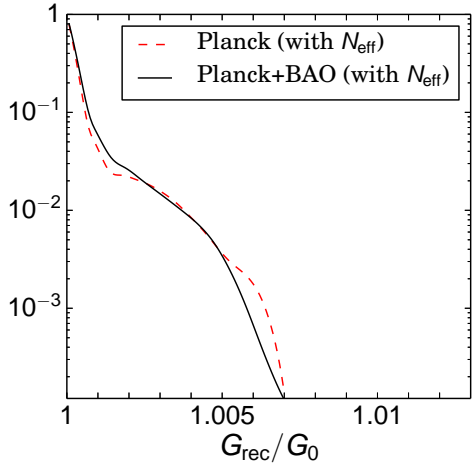}
\caption{\label{fig:Neff_BAOGphi_1d} Posterior distribution of $G_{\rm rec}/G_0$ for 
the flat models with $N_{\rm eff}$,
using the Planck data only (red dashed) or the Planck+BAO data (black).
}
\end{figure}

Also, we find that $G_{\rm rec}/G_0$ in the case including $N_{\rm eff}$ (Fig. \ref{fig:Neff_BAOGphi_1d}) is constrained as
\begin{align}
G_{\rm rec}/G_0 -1&< 2.5\times 10^{-3} \ \ (95.45\%), \\
G_{\rm rec}/G_0 -1&< 6.8\times 10^{-3} \ \ (99.99\%).
\end{align}
The right column in Table \ref{tab:table1} shows $68.27\%$ confidence limits of the cosmological parameters
in the scalar-tensor $\Lambda \rm CDM$ model including $N_{\rm eff}$.
These parameters are also still consistent with the those of the Planck results \cite{Planck}.
The limits on the ${\rm log}_{10}({\alpha_0}^2)$ and $\beta$ are summarized in Table \ref{tab:table2}.

\subsection{Choice of prior}

So far, we have performed the analysis using the flat prior on $\log\alpha_0$.
In fact, it is equally possible to perform the
analysis using the flat prior on $\alpha_0$. There seems no
preference for the choice of the prior. However, the
distributions of the
prior differ greatly depending on the choice of variable: {}from the
Jacobian due to the change of variable in the distribution function, the
uniform distribution in terms of $\alpha_0$, $P(\alpha_0)={\rm const.}$,
corresponds to a preference for large $\log \alpha_0$ in terms of $\log
\alpha_0$, $P(\log \alpha_0)\propto \exp(\log \alpha_0)$, or the uniform
$P(\log\alpha_0)$ corresponds to a preference for small $\alpha_0$ in
terms of $\alpha_0$, $P(\alpha_0)\propto 1/\alpha_0$.  In this section, we
discuss the consequence of the choice of the prior for the constraints on the
parameters. The effects of the choices of priors in anisotropic
universes are discussed in \cite{bianchi}.

 We perform the analysis using the uniform prior on $\alpha_0$. Namely, 
instead of Eq. \ref{eq:prior2}, we set
\beqa
\alpha_0  \in (0, 0.5).
\eeqa
The priors on the other parameters are the same as Eqs. (\ref{eq:prior}) and (\ref{eq:prior3}). 

\begin{figure}[ht] 
\centering\includegraphics[width=7cm,]{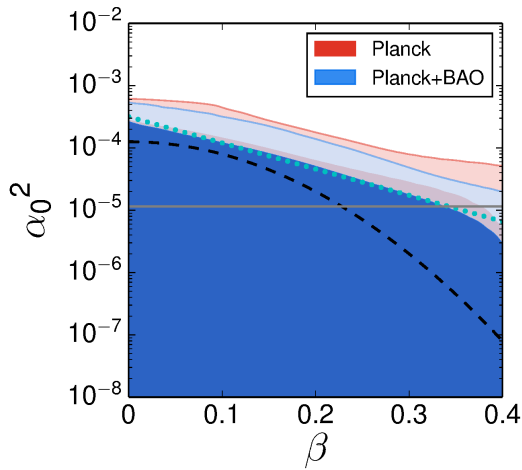}
\caption{\label{fig:BAO_contour} Contours for $95.45\%$ and $99.99\%$ confidence levels in the ${\alpha_0}^2$--$\beta$ plane
for the scalar-tensor $\Lambda {\rm CDM}$ models with the other parameters marginalized,
using the Planck+BAO data.
The cyan dotted line show the function ${\alpha_0}^2 =  10^{-3.5-4.2\beta}$.
The black dashed line and the gray solid line are the same as Fig. \ref{fig:flat_BAOcontour}.
}
\end{figure}

\begin{figure}[ht] 
\centering\includegraphics[width=7cm,]{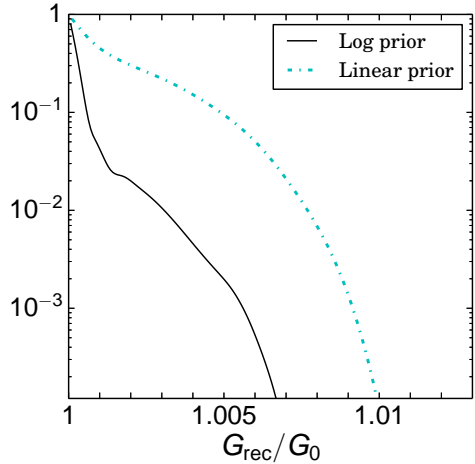}
\caption{\label{fig:Gphi_log_linear} Posterior distribution of $G_{\rm rec}/G_0$.
}
\end{figure}

In Fig. \ref{fig:BAO_contour}, the constraints  in the
$\alpha_0^2$--$\beta$  plane are shown  
for both the linear prior and logarithmic prior cases.  
The posterior distribution functions for $G_{\rm rec}/G_0$ are shown in Fig. \ref{fig:Gphi_log_linear}. 

With the linear prior, we find that $\alpha_0^2$ is constrained as
\begin{align}
\log_{10}({\alpha_0}^2) &<-3.5-4.2\beta\ \ (95.45\%), \\
\log_{10}({\alpha_0}^2) &<-3.2-4.2\beta\ \ (99.99\%),
\end{align}
and the coupling parameter $\omega$ as
\begin{align}
\omega &> 2009\ \ (95.45\%), \\
\omega &> 907\ \ (99.99\%).
\end{align}
The 95.45\% confidence limit of $\omega$ is almost the same as our previous results in \cite{ooba},
while the constraint at the 99.99\% confidence limit is strengthened.

$G_{\rm rec}/G_0$ is constrained as
\begin{align}
G_{\rm rec}/G_0 -1 &< 5.2\times 10^{-3}\ \ (95.45\%), \\
G_{\rm rec}/G_0 -1 &< 8.9\times 10^{-3}\ \ (99.99\%).
\end{align}
These are a 7\% ($95.45\%$) or 20\% ($99.99\%$) improvement over our previous results obtained
by the Planck data alone:  $G_{\rm rec}/G_0-1<5.6 \times 10^{-3} \ (11.5\times 10^{-3})$ 
at $95.45\%$ C.L. ( $99.99\%$ C.L.) \cite{ooba}.
Therefore, the statistical merit of including BAO is more significant in the linear-prior case than in the log-prior case.
We find that  the constraint on $\alpha_0^2$--$\beta$ and
the constraint on $G$ with the linear prior are more relaxed than those with the flat prior on $\log_{10}(\alpha_0)$.

\section{Summary}

We have constrained the scalar-tensor $\Lambda \rm CDM$ model from the Planck data  and the BAO data by using the MCMC method.
We have found that the present-day deviation from the Einstein gravity (${\alpha_0}^2$) is constrained as
$\log_{10}({\alpha_0}^2) < -3.9-20\beta^2$ ($95.45\%$ C.L.) and $\log_{10}({\alpha_0}^2) < -2.8-20\beta^2$ ($99.99\%$ C.L.) for $0<\beta<0.4$.
The variation of the gravitational constant is also constrained as 
$G_{\rm rec}/G_0 < 1.0019$ ($95.45\%$ C.L.) 
and $G_{\rm rec}/G_0 < 1.0055$  ($99.99\%$ C.L.). 
These constraints are improved more than 10\% compared with the results obtained by the Planck data alone.
We have also found that these constraints are not much affected by 
the inclusion of the spatial curvature or the effective number of relativistic degrees of freedom $N_{\rm eff}$. 
We have discussed the prior dependence of the analysis and found that the constraints using the flat prior on
$\alpha_0$ are slightly relaxed: 
$G_{\rm rec}/G_0 < 1.0052$ ($95.45\%$ C.L.) 
and $G_{\rm rec}/G_0 < 1.0089$  ($99.99\%$ C.L.).

\section*{Acknowledgments}
This work is in part supported by MEXT Grant-in-Aid for Scientific Research on Innovative Areas 
Nos. 15H05890 (NS and KI) and 15H05894 (TC).  
This work is also supported by 
Grant-in-Aid for Scientific Research from JSPS (Nos. 24540287 (TC), 24340048 (KI) and 25287057 (NS)),  and 
in part by Nihon University (TC).   


\appendix
\section{\label{app1}Einstein frame and the Harmonic Attractor Model}


In this appendix, we explain the details of the choice of Eq. (\ref{eq:omega}). 
We define the Einstein frame metric $\barg_{\mu\nu}$ by the conformal transformation of the form
\beqa
g_{\mu\nu}=\frac{1}{\phi}\barg_{\mu\nu}\equiv e^{2a}\barg_{\mu\nu} \,.
\label{conformal}
\eeqa
Then, Eq. (\ref{eq:action}) can be rewritten as
\beqa
S &=&\frac{1}{16\pi G_0}\int d^4x\sqrt{-\barg}\left[ \barR - \left(\omega(\phi)+\frac32\right)\frac{(\barnabla \phi)^2}{\phi^2} \right] \nonumber\\
&&+ S_{\rm m}[\psi,e^{2a}\barg_{\mu\nu}] \,.
\label{eq:action:einstein}
\eeqa
We introduce the normalized scalar field $\vp$ by
\beqa
\left(\omega(\phi)+\frac32\right)\frac{(d \phi)^2}{\phi^2}=2(d\vp)^2 \,.
\eeqa
{}From Eq. (\ref{conformal}), $\omega(\phi)$ is related to $a(\vp)$ by
\beqa
2\omega+3=\left(\frac{da}{d\vp}\right)^{-2} \,.
\label{omega}
\eeqa
Note that the extrema of $a(\vp)$ correspond to $\omega\rightarrow \infty$ (the Einstein gravity). 
Since the cosmological evolution of $\vp$ is determined 
$\ddot\vp+3\barH\dot\vp=-4\pi G_0(da/d\vp)(\bar\rho-3\barp)$ (where barred quantities are to be 
regarded as those in the Einstein frame) \cite{dn,dn2}, we can regard that $a(\vp)$ is (proportional to) 
the effective potential. We Taylor-expand $a(\vp)$ around the present time up to the quadratic order:
\beqa
a(\vp)=a_0+\alpha_0(\vp-\vp_0)+\frac12\beta(\vp-\vp_0)^2\,.
\eeqa
{}From Eq. (\ref{omega}), in terms of $\phi=e^{-2a}$, this generic choice of $a(\vp)$ 
corresponds to 
Eq. (\ref{eq:omega}).



\end{document}